# Conceptual Model for Communication

Sabah Al-Fedaghi
Computer Engineering Department
Kuwait University
P.O. Box 5969 Safat 13060 Kuwait
sabah@alfedaghi.com

Ala'a Alsaqa
Computer Engineering Department
Kuwait University
P.O. Box 5969 Safat 13060 Kuwait
eng_alaa_alsaqa@hotmail.com

Zahra'a Fadel
Computer Engineering Department
Kuwait University
P.O. Box 5969 Safat 13060 Kuwait
z_fadel@hotmail.com

*Abstract*—A variety of idealized models of communication systems exist, and all may have something in common. Starting with Shannon's communication model and ending with the OSI model, this paper presents progressively more advanced forms of modeling of communication systems by tying communication models together based on the notion of flow. The basic communication process is divided into different spheres (sources, channels, and destinations), each with its own five interior stages: receiving, processing, creating, releasing, and transferring of information. The flow of information is ontologically distinguished from the flow of physical signals; accordingly, Shannon's model, network-based OSI models, and TCP/IP are redesigned.

I. INTRODUCTION

Communication is typically defined as a process of sending and receiving. Such a communication process can be found in many disciplines, ranging from psychology and sociology to engineering, technology, and artificial intelligence. Consequently, great interest has been shown in finding an idealized communication model that provides "both general perspective and particular vantage points from which to ask questions and to interpret the raw stuff of observation" [8].

A communication *model* is an idealized systematic representation of the communication process. Such models serve as standardization tools, and they provide the means to

*1)* question and interpret actual communication systems that are diverse in their nature and purpose,

*2)* furnish order and structure to multifaceted communication events, and

*3)* lead to insights into hypothetical ideas and relationships involved in communication.

A variety of communication systems models exist, and "perhaps they all [have] something in common" [12]. Shannon's model of communication and its variations are the most common models adopted in many fields. The seven-layer OSI model is well known as a reference model for describing networks and network applications. It is a reference model for the five-layer TCP/IP model. The OSI model can also be extended to include a human perspective, as will be described in this paper.

The need for a *general* communication model can be seen in the evolution of the original Shannon's model based on efforts of engineers to find the most efficient way of transmitting electrical signals. Nevertheless, the model has been enhanced to interpret all instances of communication, that is, to organize biological communication systems along the same lines as telecommunications systems, with the notion of interactivity overcoming the linearity of the original model.

Modeling communication is an evolutionary process in which new concepts enhance and complement earlier communication models. This paper presents one more step in the evolutionary process of models with a proposal to base modeling of communication on the notion of flow. It ties communication models together through a *flow model of communication* that focuses on abstract description without involving details of the communication environment. This flow-based model contributes to building an idealized communication model through enhancing and integrating different conceptualizations of the communication process. It is different from other models in three main aspects:

- Most current communication models treat participants (e.g., nodes) in the communicative act as a send/receive system. In the flow-based model, the interior anatomy of the participants in the communication process includes stages of receiving, processing, creating, release, and transfer of information. This provides many advantages, such as the ability to identify the participant's role in communication acts. For example, the sender may be just a mere receive-and-send agent (e.g., dumb terminal), or a source (creator) of the transferred information, and so forth.

- Most current communication models do not explicitly distinguish among different types of flow (e.g., information, messages, and signals). Such a conceptualization is analogous to representing the gas, water, and electricity lines in the design of a building by one type of arrow in the design blueprint. In the flow-based model, each type of thing that flows has its own map of flow that can trigger other types of flow.

- Most current idealized communication models do not grant the channel of communication full status as a participant in the communication process. In contrast, in our model, the channel incorporates full functionality equal to that of other participants; that is, it receives, processes, creates, releases, and transfers information, as will be described.

## II. MODELS OF COMMUNICATION

Hartley [6] was the first to quantify "signals as means to convey information" through the equation $I = N \log S$, where $I$ is the amount of information each message contains, $N$ is the number of signs in a message, and $S$ is the number of different signs in the vocabulary. Shannon formalized information as reduction of uncertainty: $I = \log_2 C$, where $I$ is the amount of information each message contains, and $C$ is the number of possible choices. Shannon and Weaver [11] point out that transmission in such a model conveys physical codes. The "meaning" is taken out prior to transmission and reinstated after reception through encoding and decoding, respectively.

Shannon's model (Figure 1) has influenced all communication models. Shannon also introduced a mechanism that accounts for differences between the transmitted and received signals; this has evolved into the current feedback concept.

If such a model were applied to human communication, "effectively, the model proposes a speaker consisting only of a mind (the source) and a mouth (the transmitter), and a listener consisting only of ears (the receiver) and a mind (the destination). It therefore totally fails to reflect the many intermediate cognitive processing stages" [12]. Accordingly, cognitive communication models have expanded Shannon's model to incorporate some of these intermediate cognitive processing stages. Smith [12], as shown in Figure 2, illustrates how at least some of this intermediate processing can be represented. The model now includes three intermediate layers at either end of the transmission channel.

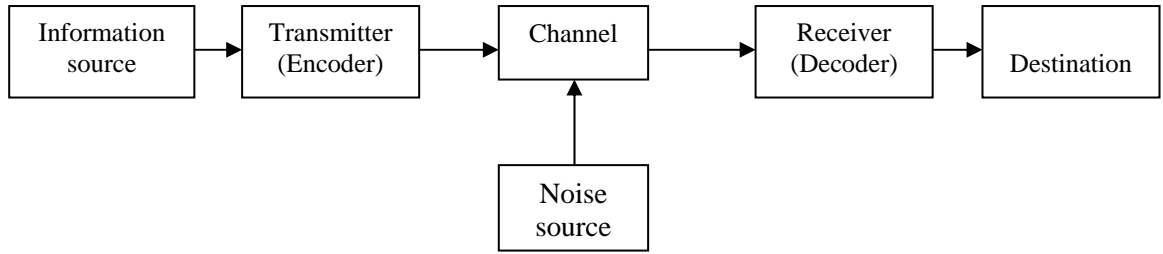

**Figure 1. Shannon's model of communication.**

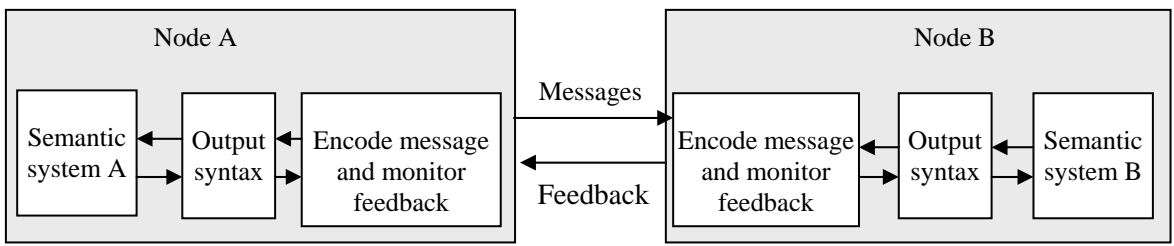

**Figure 2. Idealized communication system (modified from Smith [12] after Osgood and Sebeok [9]).**

In our flow-based perspective, Shannon's model reflects a limited view of *states* of entities being communicated, because in such a model information is observed in either *sent* or *received* states. Implicitly, the model indicates that information passes, or is *transferred*, through the presence of a channel in the model. Such a view is analogous to conceptualizing the notion of travel as *transfer* from one point and *arrival* at another point. In a more comprehensive view, the process of (air) travel includes the notions of being *received* at the travel station (airport), *processed* (e.g., luggage and passports), *released* to boarding (waiting for boarding), and actual *transfer* onto the plane. On the other end, after being *transferred*, passengers *arrive*, are *processed*, *released*, and then *transfer* (leave for hotels). Similarly, information in the communication stream is not just sent and received, but also has repeated lifecycle states: received, processed (changing its form), created (generating information from information), released (e.g., waiting for channel to be available), and transferred.

A basic claim in our communication model is that the life cycle of information in any communication system consists of iterations of stages according to a *state diagram* that will be described later. The five stages are receiving, processing, creation, release, and transfer. Life cycle refers to the "birth" of a communicative act through initiation of a flow of information directed to a certain destination and the "decay" of such an act through the seizure of flow of information. The seizure or stoppage of flow of information can occur at any point in the flow stream of information regardless of whether a destination is reached. The flow stream is successive stages of the stages described previously across different participants' boundaries.

In addition, in our flow-based perspective, Shannon's model does not reflect conceptually the ontological nature of communication. For example, it is very well known that a communication act involves information, a message (symbolic representation), and signals (e.g., physical or electronic signals). The flow of these three types of things is represented by a single arrow between the sender and receiver. Such a conceptualization is analogous to representing the gas, water, and electricity lines in the design of a building by one type of arrow in the design blueprint. Information is usually created by the sender, while noise is created in the channel. The noise is physical signals. Conceptually, this noise ought not be mixed with randomness (entropy) created at the source in some applications. Entropy is a "type of information," while noise generated in the channel comprises physical signals. Of course, noise interweaves with messages while being converted to signals for transmission purposes.

In the flow-based model, the thing that flows has its own map of flow that can trigger other types of flow. The criticisms outlined above can be applied to the next major development in modeling of communication: the OSI model.

## III. OSI MODEL

The evolution of idealized communication models evolved with the seven-layered Open Systems Interconnection (OSI) that includes many details such as authentication, routing identification, governing, data compression and decompression, and detection of errors in transmission and arranging for their correction. In this paper, we concentrate on its main feature as a model of communication.

The seven layers of the OSI model were established in 1977 by the International Organization for Standardization. It is a reference tool for understanding data communications. It represents the entire process of transmitting data from one computer to another. It divides the communications process into seven layers, as follows:

**Layer 7—Application layer:** This is the "end-user" level of communication. It is the level of pragmatic exchange between minds [12]. It is the point of origin of the message intended to be communicated by a sender, and the point of final arrival of the message as interpreted by a receiver.

**Layer 6—Presentation layer:** This is the stage where surface syntactic structure is created in outgoing messages and interpreted in incoming ones. In computer networks, this is the level at which data encryption and compression take place.

**Layer 5—Session layer:** This layer sets up, manages, and terminates, when necessary, the lower layers of the communication link. It identifies and authenticates the recipients and controls the passing of Layer 6 information downward and upward. It also synchronizes the activities of transmitting and receiving so that stations do not end up all talking at once.

**Layer 4—Transport layer:** This is where information from layers 5–7 is translated into a format compatible with the physical link. This process includes error checking and peer-to-peer transmission acknowledgment. It begins the process of message fragmentation into "packets," manages the transfer session, and, in an analogy to human communication, frequently uses "facial expressions" and "gestures" to exchange its Transport layer messages [12]. A receiving Transport layer assembles incoming messages from its transmission packets back into units that can be processed, such as words and phrases.

**Layer 3—Network layer:** This is where the transmission path is decided. This layer is needed only in large networks where there are optional routes between nodes.

**Layer 2—Data Link layer:** This is where the information is formed into transmittable signal strings utilizing such instruments as hubs and switches.

**Layer 1—Physical layer:** This is the bit-level transmission layer. It transmits the signals in a particular format characterized by connector types, cable types, voltages, and pin-outs.

Content:

Figure 3 illustrates the OSI communication system between two nodes.

The OSI model explains networking in general terms. It has been used as an educational tool and as an illustration of interactions between communication protocol suites and devices. Again, as with Shannon's model, we see that the OSI model is basically a send/receive model. The seven layers are transformations of different things that flow. To simplify, a user's information is transformed to messages that are transformed to signals; thus, the thing that flows is different along the communication chain. These characteristics are conceptually disturbing.

To complete the picture of important conceptualizations of communication, we next describe the model of Transmission Control Protocol/Internet Protocol (TCP/IP).

## IV. TCP/IP MODEL

The Internet has given rise to TCP/IP (Transmission Control Protocol/Internet Protocol) communication protocols. TCP/IP includes five layers that correspond in general to the OSI model (see Figure 4) and provides a framework for various protocols such as HTTP (which runs the World Wide Web) and FTP. The five layers of TCP/IP are described as follows:

**Application (Layer 5):** Handles everything else handled in the lower layers.

**Transport (Layer 4):** Manages all aspects of data routing and delivery, including session initiation, error control, and sequence checking.

**Internet (Layer 3):** Responsible for data addressing, transmission, and packet fragmentation and reassembly.

**Network access (Layer 2):** Specifies procedures for transmitting data across the network, including how to access the physical medium.

**Physical (Layer 1):** Covers the physical interface between a data transmission device and a transmission medium or network.

## V. THE FLOW MODEL

The flow model (FM) was introduced by Al-Fedaghi and has been used since 2006 in several applications such as description of information flow. While this section reviews the basic seeking, information security, and database access control aspects of the model to make the paper self-contained, it also presents new illustrations of the model.

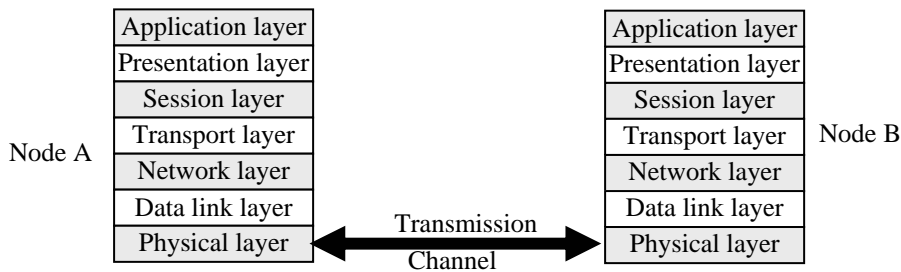

**Figure 3. Seven-layer communication system (modified from Smith [12], simplified from Purser, 1987).**

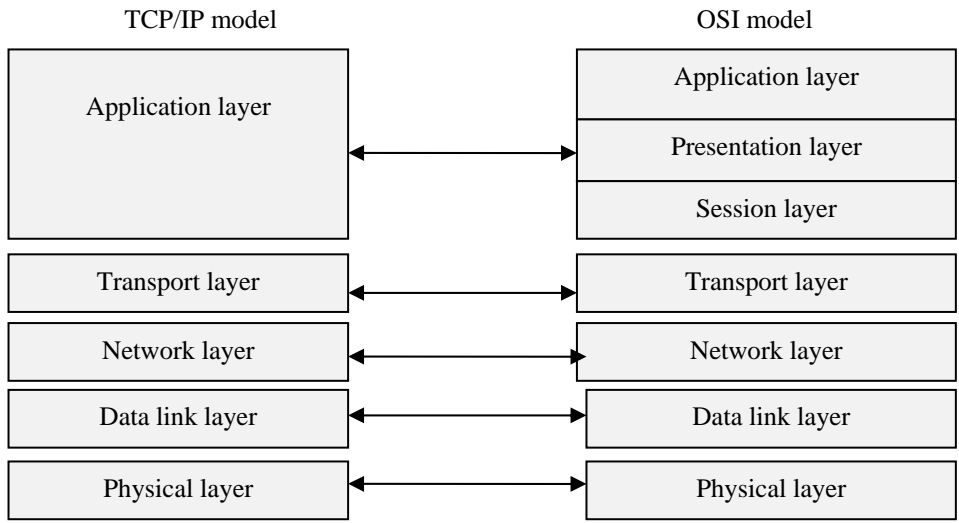

**Figure 4. TCP/IP layers mapped to the OSI layers.**

FM has a number of different components and uses a spatial assembly of these components relative to each other and to time, and it shows the links between the components that indicate the flow of items. To simplify this review of FM, we introduce flow in terms of information flow.

Information goes through a sequence of states as it moves through stages of its lifecycle, as follows:

*1)* Information is *received* (i.e., it arrives at a new sphere, similar to passengers arriving at an airport).

*2)* Information is *processed* (i.e., it is subjected to some type of process, e.g., compressed, translated, mined).

*3)* Information is *disclosed/released* (i.e., it is designated as released information, ready to move outside the current sphere, such as passengers ready to depart from an airport).

*4)* Information is *transferred* (disclosed) to another sphere (e.g., from a customer's sphere to a retailer's sphere).

*5)* Information is *created* (i.e., it is generated as a new piece of information using different methods such as data mining).

*6)* Information is *used* (i.e., it is utilized in some action, analogous to police rushing to a criminal's hideout after receiving an informant's tip). Using information indicates directing or diverting the information flow to another type of flow such as actions. We call this point a *gateway* in the flow.

*7)* Information is *stored*. Thus, it remains in a stable state without change until it is brought back to the stream of flow again.

*8)* Information is *destroyed*.

The first five states of information form the main stages of the stream of flow, as illustrated in Figure 5. When information is stored, it is in a substate because it occurs at different stages: information that is created (stored created information), processed (stored processed information), and received (stored received/row information).

The five-stage scheme can be applied to humans and to organizations. It is reusable because a copy of it is assigned to each agent or entity. Consider an information sphere that includes a small organization with two departments; it is made up of three information schemes: one for the organization at large, and one for each department. Figure 6 shows the internal information flow in such a sphere.

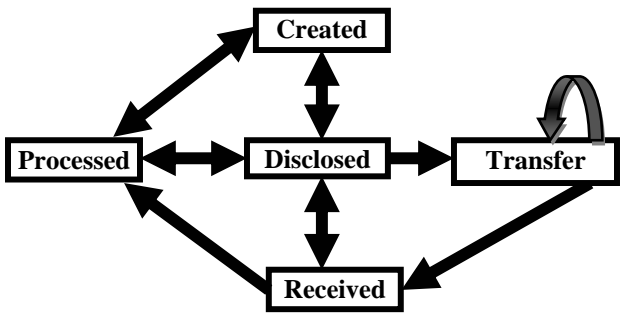

**Figure 5. Information states in FM.**

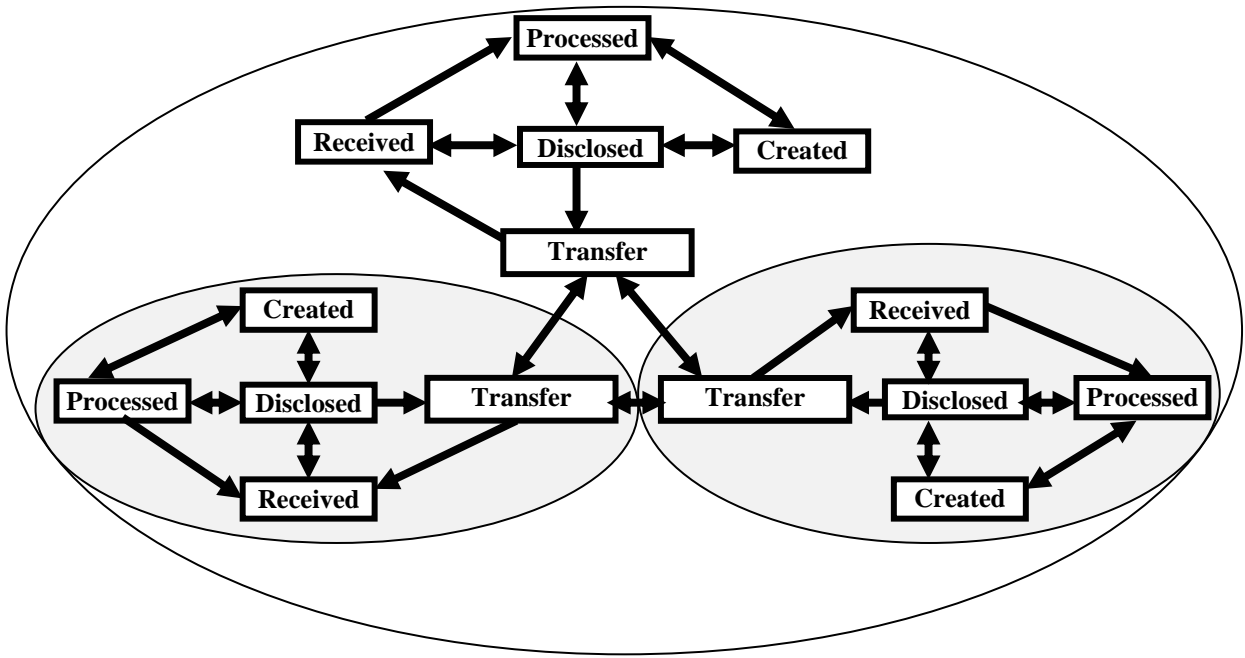

**Figure 6. Information flow within a company and its two departments.**

The five information states are the only possible "existence" patterns in the stream of information. To follow the information as it moves along different paths, we can start at any point in the stream. Suppose that information enters the processing stage, where it is subjected to some process. The following are ultimate possibilities:

*1)* It is stored.
*2)* It is destroyed.
*3)* It is disclosed and transferred to another sphere.
*4)* It is processed in such a way that it generates new information (e.g., comparing certain statistics generates that *Smith is a risk*).
6) It triggers another type of flow. For example, upon receiving patient health information, the physician takes some action such as performing medical treatment. Actions can also be received, processed, created, released, and communicated.
Notice that the arrows between Release on one hand and the stage of Received, Processed., and Created are bidirectional. This flow in opposite directions accounts for the case when it is not possible to communicate information, as in the case of a broken channel. In this case, at the Release stage, the information can be destroyed after a certain period, stored indefinitely, or returned to the releaser at the receiving, processing, or creation stages.

VI. FLOW-BASED APPROACH TO SHANNON'S MODEL

The flow model assumes that parties involved in the communication act are represented by the five components or stages: receiving, processing, creation, release, and transfer. According to Shannon, the different elements involved in communication information are source, transmitter, channel, receiver, and destination.

*Source/Transmitter*

The source produces messages to be communicated to the receiving terminal. FM extends this side of communication to highlight the "origin" of the message, whether received from outside the source, or created within the source; thus, the source can be described as creator or recipient, in addition to being a sender of the message. Such a qualification may be significant in certain circumstances (e.g., networking where communication involves a chain of two-party exchanges).

The transmitter converts the message to a signal suitable for transmission over the channel. In FM, the source has two spheres: the messages sphere and the signal sphere. Thus, this element in Shannon's model reflects the source as a processor that triggers the creation of signals that are released and transmitted.

Figure 7 shows the conceptualization of the source in FM. First, this conceptualization distinguishes explicitly between two flowthings: information and signals, thus separating the flow of information from the flow of signals.

Shannon's information theory makes a clear distinction between signals and information. In many communication systems, a signals transmission is involved only in transferring data, without the direct intention that data *conveys* information. In conventional terminology, the notion of *data* is introduced as a form of information more suitable for transmission. Looking at data from the FM point of view, data is *processed* (stage in FM) as digitally encoded information. We thus have two ways to conceptualize the relationship between information and data. If *data* is viewed as a different flowthing from information, then another sphere (besides information and signal) for data is distinguished in FM; however, in the communication context, without loss of generality, we view data as a *form* of processed information.

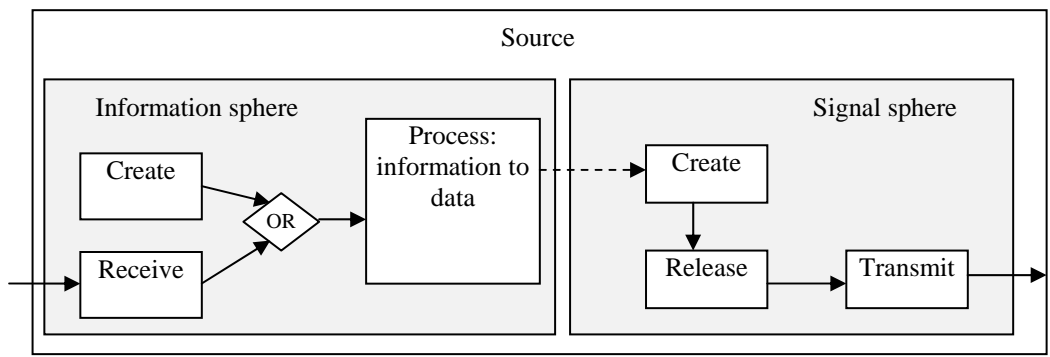

**Figure 7. Conceptualization of the source in the communication process.**

*Receiver/Destination*

Shannon's model abstracts *transmission* as a component of the source and *receiving* as a component of the destination. This is a reasonable way to look at the communication process, because "transmission" requires a deliberate act of message releasing, and "receiving" requires another deliberate act of accepting the message. Even if "transmission" is conceptualized as being in the channel proper, the notions of "transmission" and "receiving" are still decisions to be made at the source and destination, respectively. For example, it is possible that a message (e.g., an e-mail) arrives at its destination; however, the communication process is not completed if the receiver deletes it without reading it.

One objection to Shannon's model is that "the receiver is constantly being fed pieces of information with no choice in the matter—willingly or unwillingly" [7]. FM is a more suitable conceptual representation since it divides communication into two types: one under the control of the sender or receiver and one in the channel.

*Channel*

If the "transmission channel" carries the signal from its "transmitter" to a "receiver" (e.g., device), then this physical activity is different from the abstract pre/post transfer stages of "releasing" and "receiving" the message at the source and destination spheres. Furthermore, the nature of the channel is different from the "nature" of the source and the destination. Clearly, the source and the destination deal first with information, whereas the channel is a "physical sphere" that deals (in this case) with physical signals. Therefore, the *basic* thing that is flowing ("transmitted" and "received" in the source and destination, respectively) is information, while the thing that flows in the channel is only a physical signal. The message has informational form when it is *released* by the source and *prior* to channel transmission, and it returns to such a form again *after* channel transmission, when it is *received* in the destination.

The point here is that the basic flowthing at source and destination is ontologically different (e.g., a different species) from the flowthing in the channel. Figure 8 illustrates this concept through the FM's three spheres: sources, channels, and destinations. Note that the "signal" sphere on both sides has the five stages as described in Figure 7.

The stages of the physical sphere are darkened to emphasize that these are stages of flow of physical signals, whereas in the other two information spheres, the stages are in a flow of pieces of information. The channel is a flow system just as the source and the destination are. The channel certainly receives, communicates, releases, processes, and creates physical signals (e.g., noise). The difference is that the channel is solely a physical sphere; therefore, Shannon's model is really a flat (with no internal structure) partial conceptual view of the channel. Implicitly, we can deduce the following from Shannon's model:

- Creation stage exists, deduced through the concept of noise.
- The channel's receiving/releasing/transfer stages are implied by the links to source and destination.
- The processing stage of the channel can be deduced by its mere act of carrying signals.

The FM conceptualization of different spheres, each with its own stages, clarifies the conceptual picture of the flow from source to destination across the channel. The flow of information in the source never crosses between the transfer stages of the source and the channel, because information flow is ontologically distinct from physical signal flow. Note that the arrows between the source and the channel and between the channel and the destination are dotted arrows. They are triggering or *transformation* arrows and not flow arrows. The abstract entity "information" cannot simply flow to or from the physical infosphere; rather, information triggers coded events in the physical sphere and is triggered by events in that sphere. Thus, in Figure 8, the flow of information leads to the emergence of physical signals at the channel. On the other hand, flow is possible if the things that flow are of the *same kind*.

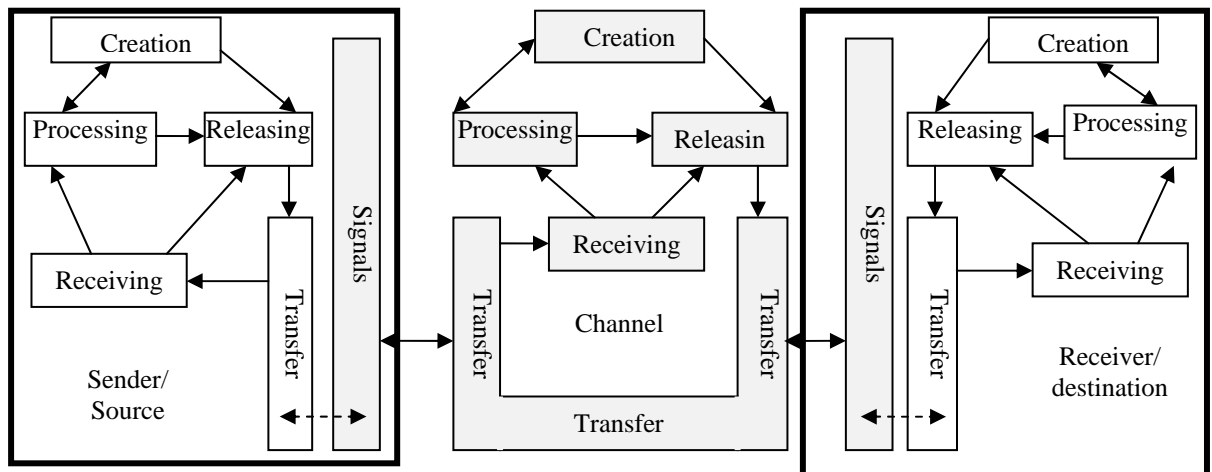

**Figure 8. FM version of Shannon's model of communication.**

Consider Figure 9, which shows channel-less communication. The two spheres can be two coupled electrical systems with current running between them. If "things that flow" were all of the same kind, we would not need channels. On the other hand, it is difficult to see such channel-less coupling between information spheres. Information is an abstract entity, so observable movement from one information sphere to another needs some type of channel.

## VII. HUMAN-MACHINE COMMUNICATION

Consider the relationship involved in the triggering mechanism between different information spheres. Figure 10 shows an FM description of information flow between two persons. First, information in the abstract information sphere of the person triggers electrical signals in the person's physiological sphere that flow from the mind/brain down the nervous system into muscles to, say, the mouth. This physiological (body) sphere can also be modeled as a five-stage sphere.

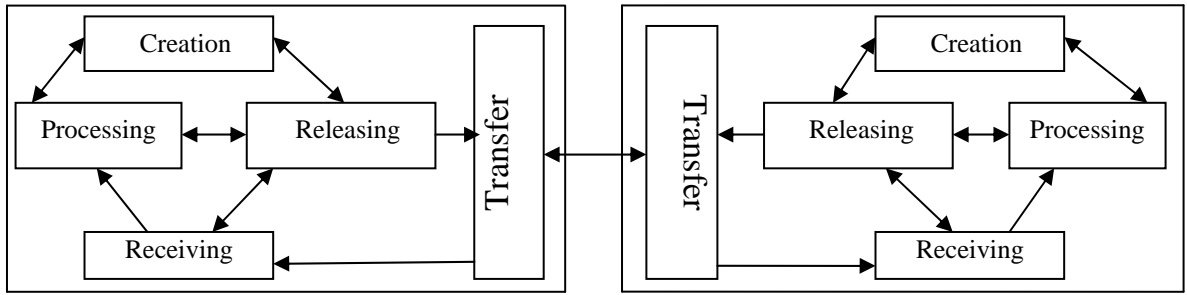

**Figure 9. FM version of direct communication between two spheres with the same types of flowing elements.**

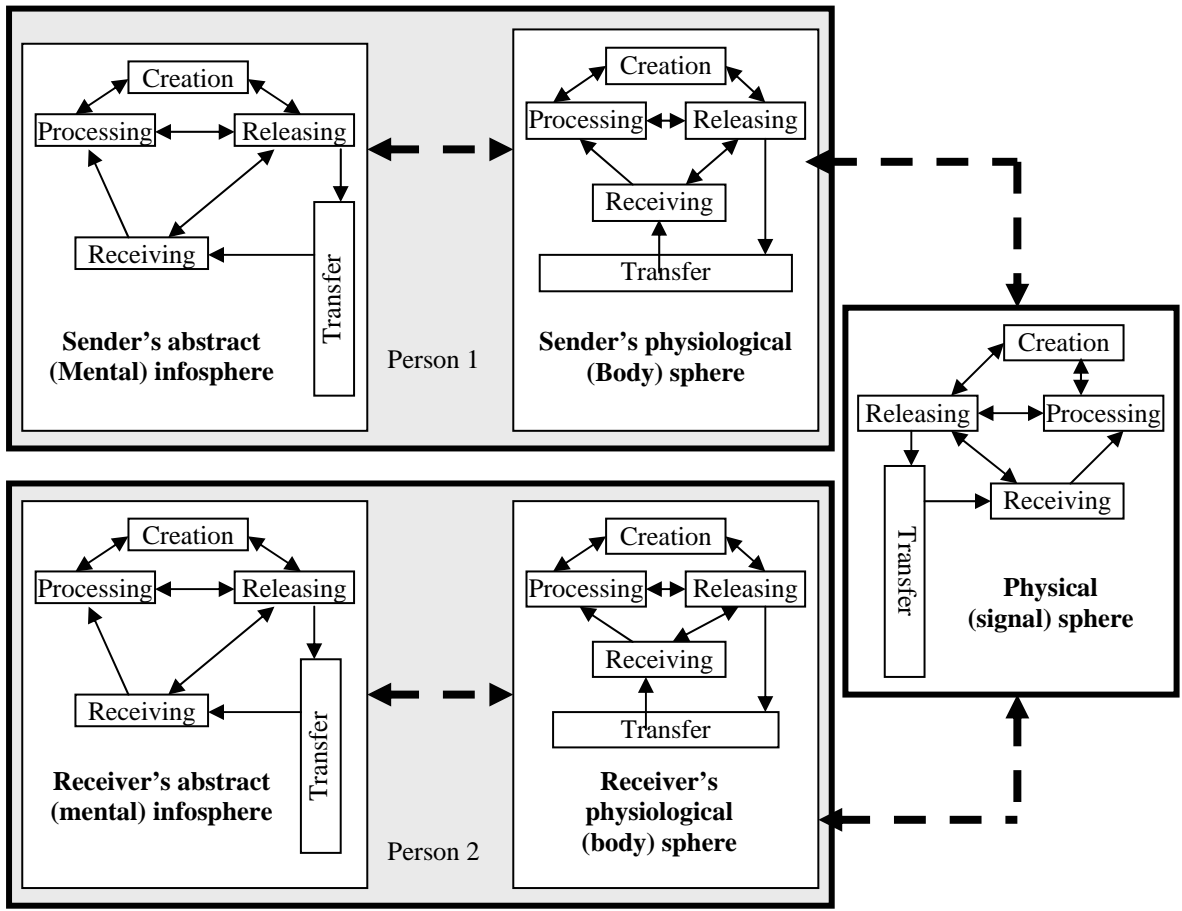

**Figure 10. FM modeling of transfer of information from one person to another.**

Notice that the model can be applied to "things that flow" or flowthings. Flowthings are things that can be received, processed, created, released, and transferred. They include information, electrical signals, materials (as in supply chains), abstract things (e.g., customer orders), and even physiological "things" (e.g., thoughts), and physical actions. Thus, the physiological system can be viewed as a communication system where electrical signals flow from the brain to, say, the mouth. The (electrical) signal in the physiological sphere triggers a sound wave signal in the physical sphere through movement of the mouth. This signal in the physical sphere reaches the ear of the receiver and triggers an electrical signal in the physiological sphere of that receiver. This physiological signal in turn reaches the information sphere in the receiver's brain.

In this scenario, three elements can be identified: the two persons and the (physical) environment. Each person has two sub-spheres: the physiological sphere (flow of electrical signals), and the informational sphere (flow of "abstract things" called information). The emergence of triggered flow appears in any stage of the sphere. For example, the information sphere may trigger the physiological sphere to *create* a signal (creation stage), or it may trigger a signal that is already *stored* (in one of the stages) in the physiological sphere (reflexes).

Osgood and Sebeok's [9] idealized communication model, shown in Figure 1, mixes the ontologies or spheres of flow between the sender and the receiver. The arrows further confuse the conceptual picture. Inside nodes (see Figure 1), the arrows seem to indicate transformation of different forms of information (semantic, syntax, encoded); however, between nodes, the arrows seem to denote flow of signals. The notations are not clear in comparison with Figure 10. The figure has five spheres with precisely declared transformations and semantics of flow.

## VIII. NETWORK MODEL

Where does the network communication model fit into this framework of communication of information? Again, examining the ISO model, we notice the following:

**Layer 7:** The application is not the application itself, although some applications may perform some of its functions.

**Layer 6:** The Presentation layer mixes the functions of spheres (sender's and receiver's) with those of channels. The sender (by implication also applied to the receiver) may prepare the message for communication, but this is different from channels processing the signal. For example, the sender may compress or encrypt the message, but such processes are different from compression and encryption of the channel.

**Layer 5**: The Session (data flow control) layer manages the lower layers of the communication link. It seems also to include functions that can be located in the sender's and channel's domains. For example, terminating the communication can be performed by the channel or by the sender.

**Layer 4:** The Transport layer includes such functions as converting address forms (e.g., eng.ku.edu into 110.10.88), checking errors, acknowledging, confirming the arrival of the entire message/signal, etc. Some of these, also, seem to be operations that can (possibly) be performed by the sender (on the message) and by the channel. This layer is said to be comparable to "human communication [that] frequently uses facial expression and gesture to exchange its Transport Layer messages. A receiving station's Transport Layer has the task of concatenating incoming messages back from their transmission packets into semantically 'processable' units such as words and phrases" [12].

**Layers 1–3:** These layers seem to be in the domain of the channel. Note that in developing a conceptual idealized communication system, we are not concerned with a particular means or technology, for example, e-mail, telephone, conversation sound, address of physical lot, 32-bit IP address, multiplexing using ports, ZIP codes, envelopes, datagrams, routers, and so on.

- **Layer 1:** This layer is the carrier of physical (electrical and mechanical) "data" stream between the sender and receiver. It is the bits flow layer. Logically, it is a *single* platform that links them.

- **Layer 2**: The data link provides synchronization for the physical level. It is the packet flow layer. Logically, it is split into two parts: the sending end and the receiving end.

- **Layer 3**: The network layer translates the destination into a network address and selects a route for messages. It is the packet preparing, assembling, and sequencing layer. In our case, we concentrate only on sender/receiver communication.

In the OSI model, the information starts at the application layer that flows down the stack with some extra information as the message flows, until it reaches the channel. It makes distinctions between lower-level data-link and transport layers and the higher-level application layers (levels 5–7). Information is to be situated at these higher levels of the model. Again, if we consider data as a form of information, information is encoded in digital data that can be processed (e.g., compressed, encoded Ethernet→fiber-optic) and transmitted as signals. The precise points of switching from information to data and then to signals are not precise in the OSI model.

Additionally, the model does not give a precise point of crossing from the sources/sender to the channel since some functions in the lower layers are mixed sender and channel functions. Adding communication information, stripping information, compression/decompression, encryption/decryption, error checking, etc. can be performed by the sender/receiver and/or by the channel in preparation for and at the end of transmission.

We propose to completely separate these spheres, as shown in Figure 11. The sender, channel, and receiver each has five stages in the FM. The dotted arrows in the figure are recognition that information/data flow is distinguished from signal flow.

Adding different stages to the channel invites different possibilities to be explored. For example, the channel is not an implicit participant in the communication act; rather, it is fully represented as a communication sphere of signals flow. It receives, processes, creates, releases, and communicates signals. Receiving and communication are the standard functions of channels in current communication models. *Creation* in the channel is manifested by noise (a type of "signal") generated in the channel. Noise is explicitly recognized in the channel. The channel may "delay" delivering the signal (e.g., traffic congestion), thus putting it in the *released* state.

## IX. OSI MODEL EXTENSION

For humans, three additional layers are introduced in the research literature [5]. To show the applicability of the flow-based approach to different generalizations of current communication models, we concentrate on the HCI (Human Computer Interaction) model as an extension of the seven-layer OSI. It is proposed as a way to facilitate discussions between HCI practitioners on one hand, and application and network developers on the other. It extends the OSI model upward in a fashion consistent with the original OSI vision.

The HCI model consists of three layers representing people's experience with the devices and services offered by technology [5]. These layers are as follows:

**Layer 3— Human Needs**: This layer "captures the essence of why a user would interact with technology; to get something done to satisfy a need" [5]. Needs include communication, acquisition of goods and knowledge, entertainment, etc.

**Layer 2— Human Performance**: This layer captures the information processing features and limitations of users. "Many [human performance capacities] are direct results of the properties of the sensory organs and the brain . . . Audio and video codices take advantage of the spatial and acoustic band pass nature of human perception" [5].

**Layer 1— Display**: This layer "represents that aspect of the hardware, software, and interfaces that a user experiences. Here at the lowest HCI layer a representation of the data is created out of signals that the human cannot understand directly (packets, bits, etc.) and that representation is displayed on a device of some sort (printer, force-feedback pointer, etc.) and used as input to Layer 9. It also works in the opposite direction to translate user output into a form that the OSI layers can understand" [5].

The three HCI layers are conceived as representing three distinct aspects of HCI that can be summarized as follows:

*1)* What a user wants to do in the abstract sense (i.e., needs).

*2)* How those needs are acted upon by the human.

*3)* The artifacts that the user employs (hardware, software, etc.).

"This common conceptual ground can be used to link applications to human needs as a function of network capabilities. The framework also helps in the discovery and localization of application performance problems and optimization opportunities" [5].

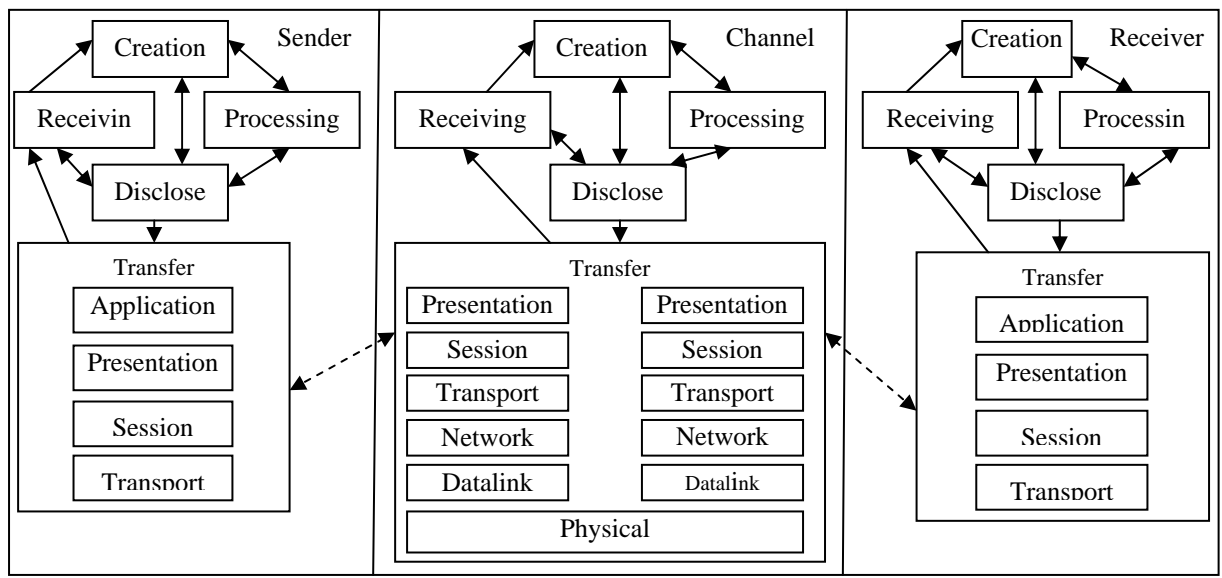

**Figure 11. Idealized FM communication model.**

Figure 12 shows a possible flow representation that involves needs, signals, actions, and information. In the FM scheme, a human being engages with several spheres according to discrete "things that flow." Similar to our use of pieces of information, we also view a need as a discrete psychological unit; thus, a sphere of needs can be assembled, as in the case of an informational sphere. Needs can be received, processed, created, released, and transferred. Received needs (desires) can be conceptualized as "planting" needs in the sense of "importing" a desire, as in the case of commercials that make a person feel a need for something (e.g., drinking a soft drink). The flow of needs is initiated in two ways:

*1)* Internal creation of needs that flow to the release and transfer stages and are manifested as desire for something.

*2)* Implanting (receiving) of needs that may proceed in the needs flow model.

In Figure 12, we assume created needs that trigger creation of (cognitive) information (e.g., a request) that triggers creation of signals that trigger user's actions (e.g., clicking on the mouse or pressing on the keyboard).

These actions are applied to peripherals; thus, actions flow in, say, a keyboard (e.g., movement of keys) that trigger the creation of signals that flow from the keyboard to the computer. This process reaches the relevant layers in the OSI model and hence proceeds in the communication stream. In such a communication scheme, we find the FM model applied uniformly at different levels: psychological, cognitive, physical signal, physical actions, etc.

### X. FLOW-BASED MODEL FOR TCP/IP MODEL

Similar to the OSI model, the flow-based approach can be applied to conceptualize the TCP/IP layers. Figure 13 shows the FM that corresponds to the five layers of TCP/IP. Similar to the process in the flow-based OSI model, information/signals flow from the sender through the channel to the recipient. Each receiver, recepient, and channel has five stages of flow. The channel receives, processes, creates, releases, and transfers signals.

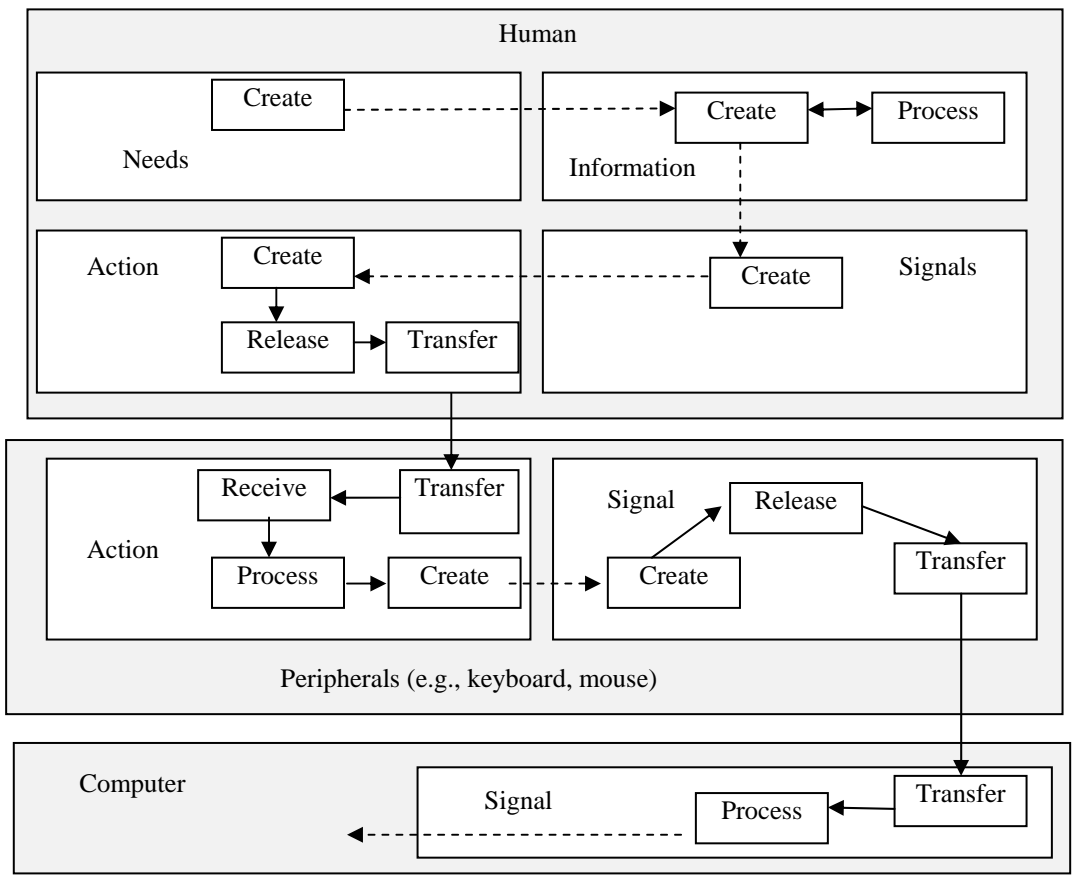

**Figure 12. Possible flows present in human-computer interaction.**

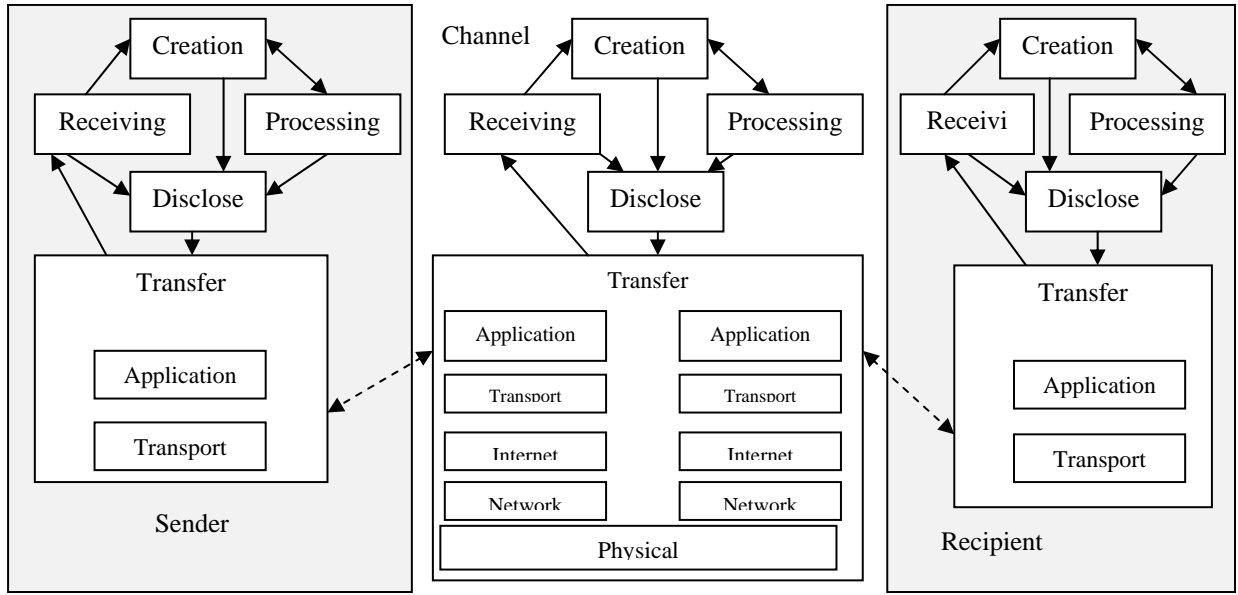

**Figure 13. FM version of TCP-IP model of communication.**

## X. CONCLUSION

The flow model enhances conceptualization of the communication process. It introduces two main points:

*1)* Instead of modeling communication spheres as closed boxes, the interior anatomy of different spheres includes five standard stages.

*2)* The communication process is modeled as the movement of "things that flow," such as information and physical signals.

These features greatly improve idealized models of communication utilized in diverse areas of application. For example, in psychology, a human information system may be modeled as a five-stage system with its own "things that flow" (e.g., thought) during participation in the communicative act.

REFERENCES

[1] S. Al-Fedaghi. "Conceptualizing effects and uses of information." Information Seeking in Context conference (ISIC 2008), Vilnius, Lithuania, September 17–20, 2008.

[2] S. Al-Fedaghi, K. Al-Saqabi and B. Thalheim. "Information stream based model for organizing security." Symposium on Requirements Engineering for Information Security, Barcelona, Spain, March 4–7, 2008.

[3] S. Al-Fedaghi. "Beyond purpose-based privacy access control." The 18th Australasian Database Conference, Ballarat, Australia, January 29–February 2, 2007.

[4] S. Al-Fedaghi. "Some aspects of personal information theory." 7th Annual IEEE Information Assurance Workshop, United States Military Academy, West Point, NY, 2006.

[5] B. Bauer and A. S. Patrick. "A human factors extension to the seven-layer OSI reference model." 2004. Retrieved from http://www.andrewpatrick.ca/OSI/10layer.html

[6] R. V. L. Hartley. "Transmission of information." Bell System Technical Journal, 7, 535-563, 1928.

[7] M. McLuhan. Understanding Media: The Extensions of Man. New York: The New American Library, 1964. Excerpt: "Interpretations with Limitations: A Critical Essay on the Mathematical Theory of Communication" at http://www.uweb.ucsb.edu/~andreabautista/

[8] C. D. Mortensen. Communication: The Study of Human Communication. New York: McGraw-Hill, 1972.

[9] C. E. Osgood and T. A. Sebeok (Eds.). Psycholinguistics: A Survey of Theory and Research Problems. Bloomington: Indiana University Press, 1965.

[10] C. E. Shannon. "The mathematical theory of communication." Bell Telephone System Journal, 27, 379-423, 1948. http://cm.bell-labs.com/cm/ms/what/shannonday/shannon1948.pdf

[11] C. E. Shannon and W. Weaver. The Mathematical Theory of Communication. Urbana: University of Illinois Press, 1949.

[12] D. J. Smith. "Shannonian communication theory and biological communication." 2003. Retrieved March 2008 from http://www.smithsrisca.demon.co.uk/shannonian-theory.html

[13] J. Bowers. "A communication model." 2006. Retrieved from http://www.jerf.org/writings/communicationEthics/node4.html

AUTHORS PROFILE

**Sabah Al-Fedaghi** holds an MS and a PhD in computer science from Northwestern University, Evanston, Illinois, and a BS in computer science from Arizona State University, Tempe. He has published papers in journals and contributed to conferences on topics in database systems, natural language processing, information systems, information privacy, information security, and information ethics. He is an associate professor in the Computer Engineering Department, Kuwait University. He previously worked as a programmer at the Kuwait Oil Company and headed the Electrical and Computer Engineering Department (1991–1994) and the Computer Engineering Department (2000–2007).

**Ala'a Al-Sakka**: Bachelor in Computer Engineering, Kuwait University, Cisco CCNA-1&2. She is Member of the Association for Computer Machinery (ACM). Her research interests include computer network, and computer and architecture.

**Zahra'a Fadel**: Bachelor in Computer Engineering, Kuwait University. She is a member of the Association for Computer Machinery (ACM). Her research interests include database systems, Computer network, and information security.